\documentstyle[aps,preprint]{revtex}
\begin{document}
\baselineskip=24pt
\setlength{\unitlength}{1cm}
\setcounter{page}{0}
\pagestyle{plain}
\noindent
\begin{center}
{\bf Analysis of chaotic motion and its shape dependence in a \\
generalized piecewise linear map }\\
\vspace{.25in}
{\bf  S.  Rajagopalan } $\footnote{ Permanent Address: Department
of Physics, Sreekrishna College, Guruvayur 680 102, Kerala, India.\\
e-mail: srgopal9999@yahoo.com}$
{\bf  and  M.  Sabir}  $\footnote{  e-mail:  msr@cusat.ac.in\ \ \
phone: 0484-540404 \ \ Fax: 0484-532495 }$\\
Department of Physics, \\ Cochin University of Science and
Technology,\\ Cochin 682 022, Kerala, India. \\
\vspace{1.25in}
{\bf Abstract}\\
\end{center}

We  analyse  the  chaotic  motion  and  its shape dependence in a
piecewise linear map  using  Fujisaka's  characteristic  function
method.  The  map is a generalization of the one introduced by R.
Artuso. Exact expressions for diffusion coefficient are  obtained
giving  previously obtained results as special cases. Fluctuation
spectrum relating to probability density function is obtained  in
a  parametric  form.  We  also  give  limiting forms of the above
quantities. Dependence of diffusion coefficient  and  probability
density function on the shape of the map is examined.
\newpage

Deterministic  diffusion  is a well known phenomenon in spatially
extended 1-D maps [1-8]. It has  been  proposed  as  a  possible
mechanism  to account for the behaviour of Josephson junctions in
the presence of microwave radiation [9] and of parametrically
driven oscillators  [10].  In  Hamiltonian  dynamics  [11]  also,
transport due to chaos is significant because of its applications
in celestial mechanics, confinement problems and so on. Recently,
some  exactly solvable models have been analysed [6-8]. The only
aim in these studies is the evaluation  of  the  exact  diffusion
coefficient  using a cycle expansion technique [12]. It is a well
known fact that the chaotic dynamics in spatially  extended  maps
has  two  complementary  aspects  -  diffusion and intermittency.
These are  related  to  the  probability  distribution  which  is
approximately  Gaussian  by  central  limit  theorem.  Fujisaka's
characteristic function method is a  useful  tool  for  analyzing
both  these aspects of stochasticity in such maps. In this brief report,
we apply the characteristic function formalism [13-14] to analyse
the chaotic motion in a generalized piecewise  linear  (GPL)  map
with  a  variable  shape.  It  is a generalization of the exactly
solvable  model  in  ref.[6]  allowing  analytical  study.  Exact
expression   for   diffusion   coefficient   and   a   parametric
representation for  the  fluctuation  spectrum  relating  to  the
probability  density  function (PDF) are obtained. Generalization
permits the study of the dependence of these  quantities  on  the
shape  of the map.
We notice that
GPL map with flat  peaks  is  more  suited  to  describe  systems
exhibiting  intermittency  in  time.  The
generalization brings the map in  ref.[6]  nearer  to  sinusoidal
maps  studied numerically in ref.[1]. A similar shape dependent
piecewise linear model has been  examined  in  ref.[2]  from  the
point of view of correlation times.

Chaos-induced diffusion systems have a general form [4,13]
\begin{equation}
X_{t+1}=X_{t} + P_{r}(X_{t}) = Y_{r}(X_{t}),\ \ \ \ \ \ \ \ \
P_{r}(X+1)=P_{r}(X)
\end{equation}
where $r$ is a control parameter. The sinusoidal map $P_{r}(X)= r
sin(2\Pi  X)$  is  an  example  [1].  After  the  decomposition
$X_{t}=N_{t}+ x_{t}$ where $N_{t}$ is the cell number measured in
which $X_{t}$ is located and $x_{t}  (0  \leq  x_{t}  <  1)$  the
distance  measured from the relative origin $X=N_{t}$, eq.(1) can
be uniquely rewritten as two dynamical laws:
\begin{equation}
N_{t+1}=N_{t} + \Delta (x_{t}),\ \  x_{t+1}=f(x_{t})
\end{equation}
Here  $\Delta  (x)$  is the jumping number defined as the largest
integer smaller than $x+P_{r}(x)$ and is free  from  $N_{t}$  and
$f(x)=x+P_{r}(x)  -  \Delta  (x)$, satisfying $0 \leq f(x) < 1$.
$f(x)$ is the reduced map of the extended map (1).

We analyse a piecewise linear map with variable shape of the type
in fig.(1). In the general case, the map consists of
linear  segments with slopes $\pm m_{i}, i=0,1,\cdots, h, m_{i} <
m_{i-1}$. For the cells on the bisector, the slope  magnitude  is
$m_{0}$.  For  the  $i^{th}$  cell  above  and below this cell on
bisector, the slope magnitude changes to $m_{i}$.
The reduced map consists of $k=4 h+3$ linear segments. For
$k$ increasing from 1 to $4 h+3$, these line segments have slopes
$m_{0}, m_{1}, m_{2}, \cdots m_{h}, -m_{h}, -m_{h-1}, \cdots 0,  -m_{1}$,
$-m_{2},  \cdots  -m_{h},  m_{h},  m_{h-1},  \cdots m_{2}, m_{1},
m_{0} $, and $m_{i} s$ satisfy the relation
\begin{equation}
\frac{3}{m_{0}} + \sum_{i=1}^{h} \frac{4}{m_{i}} = 1
\end{equation}

The  extended map can be generated from the reduced map by giving
suitable jump numbers $\Delta (x)$
(constant for a line
segment). These (from left) are $0, 1, 2, \cdots h, h, h-1, \cdots 2, 1, 0, -1,  -2,
\cdots  ,  -h,  -h, -(h-1), \cdots -2, -1, 0$. Figs.(1) and (2)
show the map and the reduced map for $h=1$.

The map  (1)  can be studied using the characteristic
function formalism [13-14]. In  this,  the  dynamics  of  $A_{t}$
governed  by  $A_{t+1}=B(x_{t}) A_{t}\ (t=0, 1, 2, \cdots )$ with
$A_{0}=1$ is studied. $B(x_{t})$ is a steady function of  $x_{t}$
which  evolves  according  to  the  chaotic map $x_{t+1}=f(x_{t})
(0\leq x_{t}  <1)$  [13].  Equivalently,  one  can  consider  the
dynamics   of   the   local   time   average  of  a  time  series
$\alpha_{t}=\frac{1}{t} \sum_{j=1}^{t} \ln B(x_{j})$ [14]. Map
(1)  can  be  treated by putting   $A_{t}=\exp   (N_{t}-N_{0});
B(x)=\exp  (\Delta  (x))$.  We  put $N_{0}=0$. Then $\alpha_{t} =
\frac{N_{t}}{t}$. The  long  time  dynamics  of  $N_{t}$  can  be
studied using Fujisaka's characteristic function
\begin{equation}
\lambda_{q} = \frac{1}{q} \lim _{t\rightarrow \infty} \frac{1}{t}
\ln [ \langle \exp (q N_{t}) \rangle ]
\end{equation}
$\langle  \exp  (q  N_{t}) \rangle $ is the average over a steady
ensemble and is the q-order moment of  $\exp  (N_{t})$.  One  can
expand  $\lambda_{q}$  in  the series of cumulants. The expansion
converges for $|q| < c$, $c$ being the  convergence  radius.
In this case, $\lambda_{q}$ can be approximated as
\begin{eqnarray}
\lambda_{q} &=& \lambda_{0} + D\ q \\
\lambda_{0} &=& \alpha_{\infty} = \lim_{t \rightarrow \infty}
\frac{N_{t}}{t} \nonumber
\end{eqnarray}
where  $\lambda_{0}$  is the drift velocity. $D$ is the diffusion
coefficient given by
\begin{equation}
\sigma_{t} = \langle (N_{t}-\lambda_{0}t)^{2}\rangle \approx 2\
D\ t
\end{equation}
for large values of $t$. $\sigma_{t}$ is the variance of $N_{t}$.
Asymptotic  PDF of $\alpha_{t}$ has a Gaussian component (central
limit theorem) and a non-Gaussian component. For $|q| << c$,  the
moment  $<\exp  (  q\  N_{t})>$  is  determined  by  the Gaussian
component (diffusion branch of $q$) and for $|q|  >>  c$,  it  is
determined by the non-Gaussian component (intermittency branch of
$q$).
The  PDF   $\rho_{t}(\alpha  )$ that $\alpha_{t}$
takes values between  $\alpha$  and  $\alpha+d\alpha$
can be obtained as
\begin{equation}
\rho_{t}(\alpha ) \sim \exp[ -\sigma (\alpha )t ]
\end{equation}
$\sigma    (\alpha    )$    being   the   fluctuation   spectrum,
$\rho_{t}(\alpha ) \rightarrow \delta  (\alpha  -\alpha_{\infty})$
as  $t  \rightarrow  \infty$. $\rho_{t}(\alpha )$ can be obtained
from  $\lambda_{q}$  in  parametric  form  using   the   Legendre
transform
\begin{eqnarray}
\alpha &=& \frac{d}{dq}(q\ \lambda_{q}) \nonumber \\
\sigma(\alpha ) &=& q^2 \frac{d}{dq}\lambda_{q}
\end{eqnarray}

We  first  consider  the  case  with $h=1$. The
reduced map consists of $7$ line segments with slopes (from left)
$m_{0}, m_{1},  -m_{1},  -m_{0},  -m_{1},  m_{1},  m_{0}$.  These
satisfy  (3).  The Frobenius-Perron operator ${\bf H}$ is defined
by
\begin{equation}
{\bf H} G(x)= \sum_{k=1}^{7} \frac{G(y_{k})}{|f'(y_{k})|} =
\sum_{k=1}^{7} \frac{G(y_{k})}{|m_{k}|}
\end{equation}
where  $y_{k}$  is  the  $k^{th}$  solution  of  $f(y_{k})=x$ and
$f'(x)=\frac{d}{dx}f(x)$. $m_{k}$ is the slope  of  the  $k^{th}$
line  segment  of  the  reduced  map. From eq.(9) we note that the
invariant density $p^{*}(x)$ is  uniform  ($p^{*}(x)=1$)  in  the
interval   $0\leq   x\leq  1\  [{\bf  H}p^{*}(x)=p^{*}(x)]$.  The
Lyapunov exponent $\lambda$ can be obtained as
\begin{equation}
\lambda = \frac{3}{m_{0}} \ln (m_{0}) + \frac{4}{m_{1}} \ln (m_{1})
\end{equation}
Since  $m_{i}  >1$,  we  note that $\lambda >0$ and therefore the
reduced map is always chaotic. Characteristic function  $\lambda_{q}$
can be evaluated [13] using the linear operator defined by Mori et al [15]
\begin{equation}
\widehat{{\bf H}} F(x)= \frac{1}{p^{*}(x)} {\bf H} [ p^{*}(x) F(x) ]
\end{equation}
\begin{equation}
\lambda_{q}= \frac{1}{q} \lim_{t\rightarrow \infty} \frac{1}{t}
\ln \langle \ e^{q\Delta x}\
\underbrace{ \widehat{{\bf H}}\ e^{q\Delta x}\ \widehat{{\bf H}}\
e^{q\Delta x}\ \cdots
\widehat{{\bf H}}\ e^{q\Delta x} }_{t-1}\ \ \rangle
\end{equation}
For our model ${\bf H}=\widehat{{\bf H}}$. $\Delta  (x)$ which
are constant over a line
segment are (from left) $0,+1,+1, 0,-1,-1,0$. Hence  we  get
from Eqs.(9), (11) and (12)
\begin{eqnarray}
\widehat{{\bf H}}\ e^{q\Delta x} &=&\frac{3}{m_{0}}+ \frac{2}
{m_{1}}\ e^{q}
+\frac{2}{m_{1}}\ e^{-q} \nonumber \\
&=& \frac{3}{m_{0}} + \frac{4}{m_{1}}\cosh (q)
\end{eqnarray}
\begin{equation}
\lambda_{q}=\frac{1}{q} \ln [ \frac{3}{m_{0}} + \frac{4}{m_{1}}\cosh (q) ]
\end{equation}

The result can be generalized for integer values of $h$.
Again,  slopes  satisfy relation (3). Frobenius-Perron
operator again leads to uniform invariant  density  $p^{*}(x)=1$.
Lyapunov exponent $\lambda$ and characteristic function $\lambda_{q}$ are
given by
\begin{equation}
\lambda = \frac{3}{m_{0}} \ln (m_{0}) + \sum_{i=1}^{h}
\frac{4}{m_{i}} \ln (m_{i})
\end{equation}
\begin{equation}
\lambda_{q}=\frac{1}{q} \ln [ \frac{3}{m_{0}} + \sum_{i=1}^{h}
\frac{4}{m_{i}}\cosh (i\ q) ]
\end{equation}
Map is fully chaotic since $\lambda >0$. The drift velocity $\lambda_{0}=0$,
always. The diffusion coefficient $D$ is
\begin{equation}
D=\lim_{q\rightarrow 0} \frac{d}{dq}\lambda_{q}= \sum_{i=1}^{h}
\frac{2i^2}{m_{i}}
\end{equation}
The fluctuation   spectrum  $\sigma(\alpha  )$  can  be  got  in  the
parametric form using Eq.(8)
\begin{eqnarray}
\alpha &=& \frac{ \sum_{i=1}^{h} \frac{4i}{m_{i}} \sinh (iq) }
{ \frac{3}{m_{0}}+\sum_{i=1}^{h} \frac{4}{m_{i}} \cosh (iq) }
\nonumber \\
\sigma (\alpha ) &=& q \frac{ \sum_{i=1}^{h} \frac{4i}{m_{i}}
\sinh (iq) } { \frac{3}{m_{0}}+\sum_{i=1}^{h}\frac{4}{m_{i}}
\cosh (iq)}
-\ln [\frac{3}{m_{0}}+ \sum_{i=1}^{h}\frac{4}{m_{i}} \cosh (iq) ]
\end{eqnarray}
$q=0$  gives  $\alpha  =0;\  \sigma  (\alpha  )=0$. If $+q$ gives
$+\alpha$, $-q$ will  give  $-\alpha$  without  changing  $\sigma
(\alpha  )$.  It can also be noted that maximum value of $\alpha$
is obtained by putting $q \rightarrow \infty$. We have
\begin{equation}
\alpha_{max}=h \  \  \  \sigma(\alpha_{max})= \ln (\frac{m_{h}}{2})
\end{equation}

In  the  special  case  when  all $m_{i}$'s are equal (=$m_{0}$),
eq.(17) can be summed to obtain a closed form expression for $D$.
In this case, eq.(3) gives $m_{0}=3+4h$.
\begin{equation}
D=\frac{h(h+1)(2h+1)}{3(4h+3)}
\end{equation}
With $h+1=\beta$,
\begin{equation}
D=\frac{(\beta-1)\beta(2\beta-1)}{3(4\beta-1)}
\end{equation}
which  reduces  to  $D=2/7$ for $\beta=2$ giving results
obtained previously in ref.(6).

Closed  form expression can  be got also for the special case
$m_{i}/m_{i-1}=r=$ a constant, $0<r<1$. Then $m_{i}=m_{0}r^{i}$.
From eq.(3)
\begin{equation}
m_{0}=3+\frac{4(1-r^{h})}{r^{h}(1-r)}
\end{equation}
For  every  $r$  between 0 and 1, the model becomes an exactly
solvable case. From eq.(17)
\begin{equation}
D=\frac{ 2[h^{2}+(1-2h-2h^{2})r+(h+1)^{2} r^{2}-r^{h+1}-r^{h+2}]}
{[3r^{h}(1-r)+4(1-r^{h})](1-r)^{2}}
\end{equation}

Limiting  forms  of  the  above  quantities can be obtained for a
constant $h$ as the peak shape becomes maximum flat. These can be
arrived at by taking limit $r\rightarrow  0$.  From  eq.(23),
$D$ behaves like
\begin{equation}
D=\frac{h^{2}}{2}
\end{equation}
The above limit can also be obtained by
putting $m_{i}\rightarrow \infty (i=0, \cdots  h-1)$  and  $m_{h}
\rightarrow 4$. Applying this we get the following limits
\begin{eqnarray}
\lim_{r\rightarrow 0} \lambda_{q}&=& \frac{1}{q} \ln [\cosh (qh)] \\
\lim_{r\rightarrow 0} \alpha &=& h \tanh (qh)
\end{eqnarray}
\begin{equation}
\lim_{r\rightarrow 0} \sigma(\alpha ) =
\ln [\frac{h^{2}-\alpha^{2}}{h^{2}} ]^{1/2}
[\frac{h+\alpha}{h-\alpha} ]^{\alpha /2h}
\end{equation}

In fig.(3) we plot diffusion coefficient versus $r$ for $h=2$. It
can  be  observed  that $D$ increases with increasing flatness of
the peak shape. $D$ varies from 0.9 to 2 when $r$ is varied  from
1  to  0. Increasing $h$, keeping $m_{i}=m_{0} (i=1, 2 \cdots h)$
appears to have more influence on increasing $D$. This is because
$D=2/7=0.29$ for $h=1$,  whereas  it  goes  to  0.909  for  $h=3$
(eq.(20).

The  probability  distribution function for $N_{t}$, the distance
from the origin, can be obtained using the  fluctuation  spectrum
$\sigma (\alpha)$. From eq.(7), we have
\begin{equation}
\rho _{t}(N) \sim \frac{1}{t} \exp [ -\sigma (\frac{N}{t})\ t]
\end{equation}
$\rho  _{t}(N)$  being  the PDF that $N_{t}$ takes values between
$N$ and $N+dN$. This PDF is  approximately  Gaussian  by  central
limit  theorem. In the exactly normal case $\lambda_{q}$ is given
by eq.(5) and $\sigma (\alpha )$ takes the form
\begin{equation}
\sigma(\alpha ) = \frac{(\alpha -\lambda_{0})^{2}}{4\ D}
\end{equation}
with
\begin{equation}
\alpha = \lambda_{0} + 2\ D\ q
\end{equation}
For the present model $\lambda_{0}=0$. In fig.(4) we plot $\sigma
(\alpha  )$ for different maps and compare with the Gaussian form
given in eq.(29). For  a  constant  $h$,  non-Gaussian  character
increases  with  increasing  flatness  of the map. But, as in the
case of diffusion coefficient, increasing $h$ has more  influence
in   producing   non-Gaussian   character   of   the  PDF.

To conclude, analysis  of  the  PDF
with  fluctuation spectrum brings out that intermittency and
non-Gaussian character
of the PDF increases with increasing peak height and flatness  of
the  map,  height  exercising more effect. This is important when
one  selects  models  for  describing  experiments  relating   to
diffusion. For example, systems exhibiting chaotic motion similar
to Brownian motion should have a Gaussian distribution. Maps with
linear segments having constant slope and minimum peak height are
useful  in  cases  like this. Maps with greater height with peaks
becoming more flat will be best suited  in  describing  diffusion
systems  showing  intermittency  in  time. With flatness becoming
maximum, diffusion coefficient  behaves  like  $\frac{h^{2}}{2}$,
$h$  being  the  peak  height.  Corresponding  limiting forms for
characteristic  function  and  fluctuation  spectrum   are   also
obtained.  The  limiting  form  of  fluctuation spectrum is quite
different from the Gaussian form  following  from  central  limit
theorem.

We are attempting a generalization of maps with fractional heights
given in ref.(7)
along similar lines. This work will be reported elsewhere.

We are grateful to University Grants Commission, India, for
providing financial assistance through DSA and COSIST schemes.

\newpage
\section*{Figure Captions}
\begin{itemize}
\item  {Fig.\ 1.\ Generalised Piecewise Linear (GPL) map with a
       variable shape with $h=1$. On both axes, units are arbitrary.}
\item  {Fig.\ 2.\ Reduced map of GPL map in fig.(1). On both axes,
        units are arbitrary.}
\item  {Fig.\ 3.\ Variation of Diffusion coefficient $D$ with
       $r$ for $h=2$. On both axes, units are arbitrary.}
\item  {Fig.\ 4.\ Fluctuation Spectrum $\sigma (\alpha )$ versus
       $\alpha$ for different cases. Solid lines represent actual
       $\sigma (\alpha )$ while dotted lines give corresponding
       Gaussian forms. (a) $h=1,\ m_{0}=\ m_{1}=\ 7$\
       (b) $h=1,\ m_{0}=\ 100,\ m_{1}=\ 4.1237$\
       (c) $h=2,\ m_{0}=\ m_{1}=\ m_{2}=\ 11$. On both axes, units
       are arbitrary.}
\end{itemize}
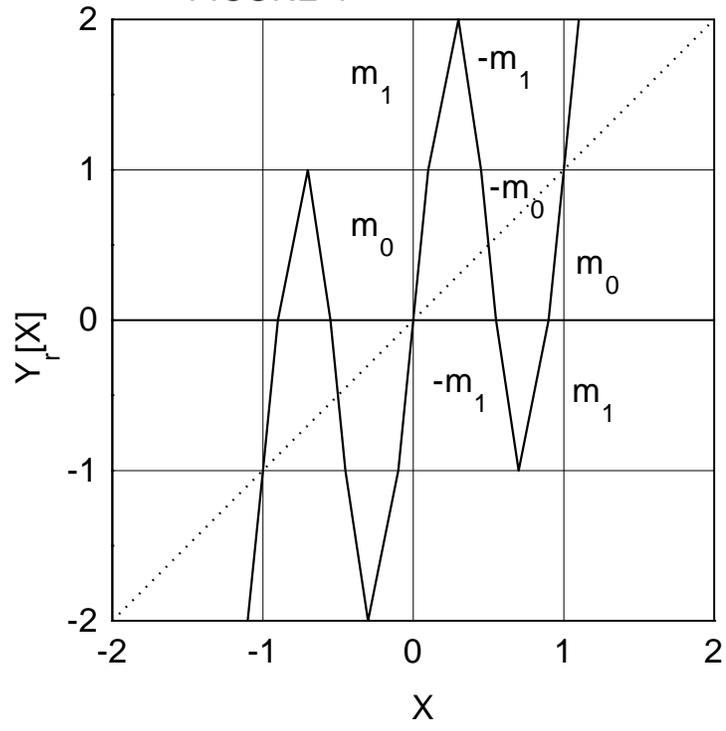
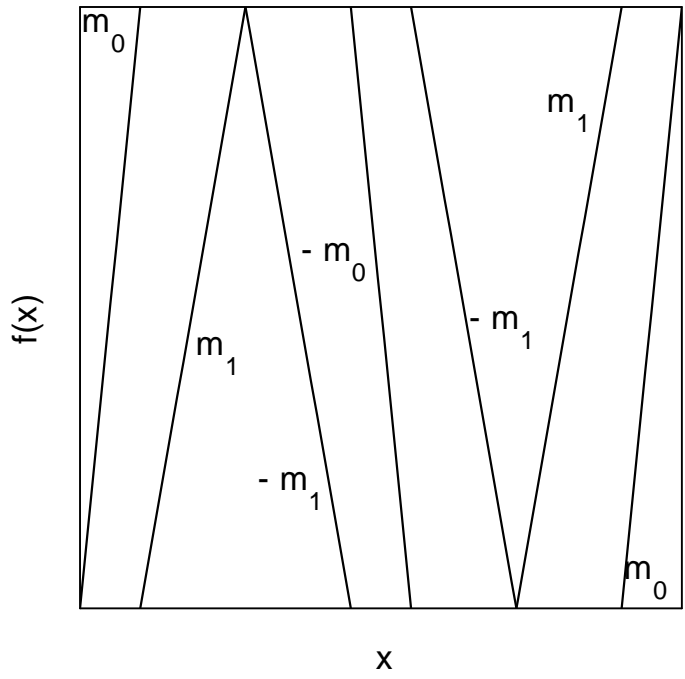
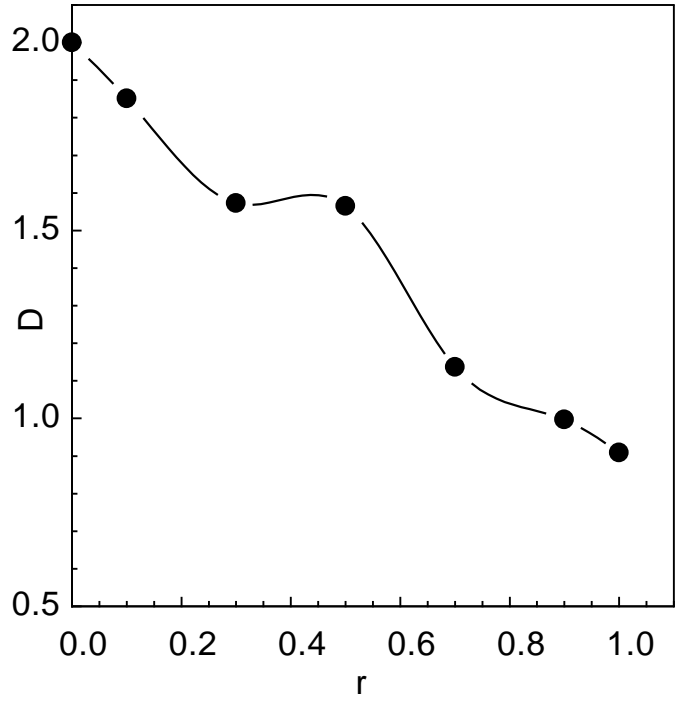
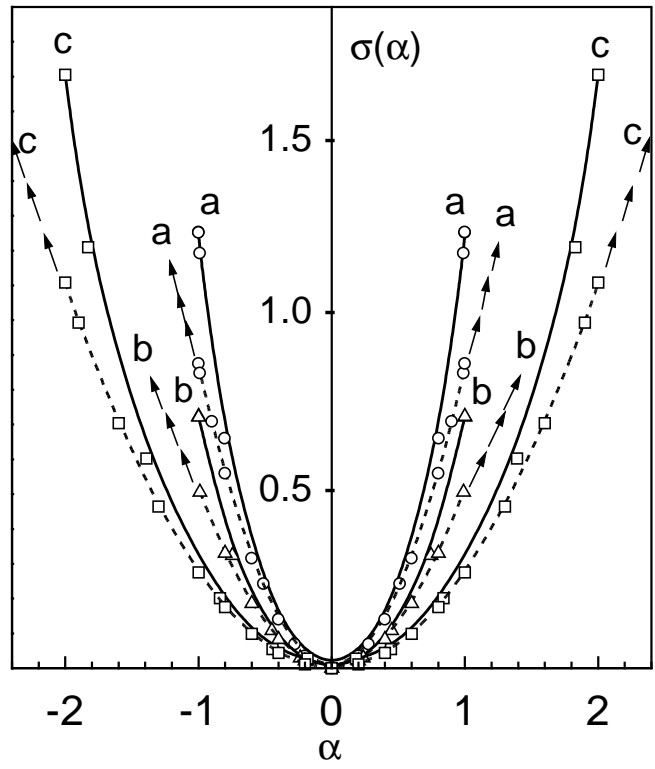
\end{document}